\documentclass[aps,prl,twocolumn,amsmath,amssymb]{revtex4}
\usepackage{graphicx,epsfig,epsf}
\usepackage{dcolumn}
\usepackage{bm}

\begin{document}
\newcommand{\ri}{{\rm i}}
\newcommand{\re}{{\rm e}}
\newcommand{\bx}{{\bf x}}
\newcommand{\bq}{{\bf q}}
\newcommand{\bu}{{\bf u}}
\newcommand{\bU}{{\bf U}}
\newcommand{\bd}{{\bf d}}
\newcommand{\be}{{\bf e}}
\newcommand{\br}{{\bf r}}
\newcommand{\bk}{{\bf k}}
\newcommand{\bE}{{\bf E}}
\newcommand{\bI}{{\bf I}}
\newcommand{\bR}{{\bf R}}
\newcommand{\bJ}{{\bf J}}
\newcommand{\bC}{{\bf C}}
\newcommand{\cL}{{\cal L}}
\def\Jp#1{J_+^{(#1)}}
\def\Jm#1{J_-^{(#1)}}
\newcommand{\bZero}{{\bf 0}}
\newcommand{\bM}{{\bf M}}
\newcommand{\bn}{{\bf n}}
\newcommand{\bs}{{\bf s}}
\newcommand{\tbs}{\tilde{\bf s}}
\newcommand{\rSi}{{\rm Si}}
\newcommand{\beps}{\mbox{\boldmath{$\epsilon$}}}
\newcommand{\rg}{{\rm g}}
\newcommand{\tr}{{\rm tr}}
\newcommand{\xmax}{x_{\rm max}}
\newcommand{\ra}{{\rm a}}
\newcommand{\rx}{{\rm x}}
\newcommand{\rs}{{\rm s}}
\newcommand{\rP}{{\rm P}}
\newcommand{\up}{\uparrow}
\newcommand{\down}{\downarrow}
\newcommand{\hc}{H_{\rm cond}}
\newcommand{\kb}{k_{\rm B}}
\newcommand{\cI}{{\cal I}}
\newcommand{\tit}{\tilde{t}}
\newcommand{\cE}{{\cal E}}
\newcommand{\cC}{{\cal C}}
\newcommand{\Ubs}{U_{\rm BS}}
\newcommand{\qq}{{\bf ???}}
\newcommand*{\etal}{\textit{et al.}}
\def\vec#1{\mathbf{#1}}
\def\ket#1{|#1\rangle}
\def\bra#1{\langle#1|}
\def\keps{\mathbf{k}\boldsymbol{\varepsilon}}
\def\dm{\boldsymbol{\wp}}
\def\CG#1#2#3#4#5#6{C{\small \begin{array}{ccc}{#1}&{#3}&{#5}\\{#2}&{#4}&{#6}\end{array}}}
\def\cLL#1#2#3#4#5#6#7#8{L{\small \begin{array}{cccc}{#1}&{#3}&{#5}&{#7}\\{#2}&{#4}&{#6}&#8\end{array}}}
\def\CGprim#1#2#3#4#5#6{C^{{#1}\,{#3}\,{#5}}_{{#2}\,{#4}\,{#6}}}
\def\CLLprim#1#2#3#4#5#6#7#8{{\cal L}_1^{{#1}\,{#3}\,{#5}\,{#7}}_{{#2}\,{#4}\,{#6}\,{#8}}}

\sloppy

\title{Coherently enhanced measurements in classical mechanics} 
\author{Daniel Braun$^{1,2}$ and Sandu Popescu$^3$}
\affiliation{$^{1,2}$ Laboratoire de Physique Th\'eorique, IRSAMC, UMR 5152
  du CNRS, 
  Universit\'e Paul Sabatier, Toulouse, FRANCE}
\affiliation{$^{3}$ H.~H.~Wills Physics Laboratory, University of Bristol,
  Tyndall Avenue, Bristol, BS8 1TL, United Kingdom} 
\begin{abstract}
We show that the recently discovered quantum-enhanced measurement
protocol 
of coherent averaging \cite{Braun11} that is capable of achieving
Heisenberg-limited sensitivity without using entanglement, has a
classical analogue.  The classical protocol
uses $N$ harmonic oscillators coupled to a central
oscillator and one measures the signal from the latter. We propose an
application to the measurement of very weak 
interactions, and, in particular, a novel route to measuring the
gravitational constant with enhanced precision.
\end{abstract}
\maketitle

A common practice for
increasing the signal to noise ratio in a measurement is to
measure identically prepared systems $N$ times and
average the measurement results.  This typically leads  to a scaling of the
sensitivity 
(i.e.~the smallest resolvable change in a parameter that we want to measure)
as $1/\sqrt{N}$, a
scaling that is known in quantum measurement theory as the ``shot
noise limit'' or the ``standard quantum
limit'' (SQL), even though there is nothing genuinely
quantum about this scaling:  It holds 
whenever the central limit theorem applies.  

With the rise of
quantum information theory, the exciting possibility that the $1/\sqrt{N}$
behavior might be improved upon has received large attention \cite{Caves81,Braunstein94,Giovannetti04,leibfried_toward_2004,giovannetti_quantum_2006,Budker07,Goda08,Nagata07,Higgins07,Taylor12,luis_nonlinear_2004,beltran_breaking_2005,roy_exponentially_2006,luis_quantum_2007,rey_quantum-limited_2007,choi_bose-einstein_2008,napolitano_interaction-based_2011,boixo_generalized_2007,Boixo08,Boixo08.2,Paris09,giovannetti_advances_2011}.
It was shown 
that if one puts the $N$ systems into an entangled state, a scaling as $1/N$
can be achieved, known as the ``Heisenberg limit'' (HL)
\cite{giovannetti_quantum_2006}.  Examples include the 
use of NOON states in a
Mach-Zehnder interferometer \cite{Bollinger96,Lee02}, or
squeezed spin states  for magnetometers
based on atomic vapors \cite{Massar03}. Unfortunately,
the entangled states required in 
these schemes are very unstable and prone to decoherence.  Experiments with
NOON states  
showing a slight improvement over the SQL have not 
surpassed yet the stage of more than a few entangled photons 
\cite{leibfried_toward_2004,nagata_beating_2007}. In fact, the
situation here is
much more unfavorable than even for a quantum computer: whereas for the
latter it should be enough to fully control a few hundred to a few thousand
qubits in order to outperform existing classical computers for specific
tasks such as 
factoring \cite{Shor95} or data fitting \cite{Wiebe12}, classical
experiments such as LIGO have already sensitivities of 
the order $10^{-22}/\sqrt{\rm Hz}$
\cite{goda_quantum-enhanced_2008,LIGO2009}. 
To compete with such 
performance using an etangled state of $N$ particles, one would 
have to entangle a macroscopic number of particles (or create a NOON state
with a macroscopic number of photons), which seems
out of reach considering the experimental
difficulties of creating a NOON state with just 4 photons.  Also, from
theoretical grounds, it has become clear that for NOON states the
slightest amount of 
decoherence leads back to the SQL scaling for sufficiently large $N$
\cite{huelga_improvement_1997,kolstrokodynacuteski_phase_2010,escher_general_2011}.
For niche applications, such in biological systems that require low
intensities, these methods may nevertheless be interesting
\cite{Taylor12}.\\

The commonly held believe
that entanglement is necessary for reaching the HL is based on propagation
of the quantum mechanical state of $N$ distinguishable particles with
a very simple hamiltonian: 
$H=\sum_{i=1}^N h_i(x)$, where $h_i(x)$ is a 
single particle hamiltonian \cite{giovannetti_quantum_2006}. No
interactions between  
different particles are considered.  Recently it was shown that a
system with $k$-body interactions offers a scaling of the sensitivity
as  
$\Delta^2_{\Psi}\,J\propto 1/N^{k-1/2}$ without initial entanglement (and $
1/N^{k}$ with initial entanglement
\cite{luis_nonlinear_2004,beltran_breaking_2005,roy_exponentially_2006,luis_quantum_2007,rey_quantum-limited_2007,choi_bose-einstein_2008,napolitano_interaction-based_2011,boixo_generalized_2007}),
even though interactions themselves will ultimately have to scale down
with $N$ if the total energy is to remain an extensive quantity.
The $k$-body interactions typically 
lead to squeezed states and ressemble in this respect the earliest examples of
quantum-enhanced measurements that proposed the use of squeezed light
\cite{caves_measurement_1980,caves_quantum-mechanical_1981}.  
With indistinguishable particles, as obtained naturally e.g.~from a
Bose-Einstein condensate, one may avoid entangling the particles as
well for surpassing the SQL limit, even though
the definition of entanglement is more tricky in this case \cite{Benatti13}. 

Another possibility is to have $N$ distinguishable systems
interact with a $N+1$st system  and read
out the latter
\cite{braun_heisenberg-limited_2011,braun_decoherence-enhanced_2009}.
This method has 
the advantage that the total system needs to accomodate only $N$ interaction
terms, such that at least in principle the interaction itself may be
independent of $N$. Furthermore, the scaling with $N$ appears to be
stable under 
local decoherence, and even decoherence itself can be used as a
signal, if the $N+1$st system is an environment.  The effect can be
understood as ``coherent averaging'': a phase accumulates  in the
state of the $N+1$st
system from the
interaction with the $N$ other systems.  No entanglement is
needed. These properties make one wonder whether a classical
analogue of this mechanism exists.  In the present Letter we
show 
that this indeed true.  More specifically, we show that there is a phase
accumulation mechanism in the classical motion of a central harmonic
oscillator interacting with $N$ other  harmonic oscillators
that is completely analogous to the quantum mechanical scenario. The
phase accumulation allows one 
to achieve a sensitivity that scales as $1/N$, just as in the quantum case with
Heisenberg-limited sensitivity, even though there is of course nothing
quantum.  Thus, while the found sensitivity is, contrary to the
quantum case, not the expression of a generalized Heisenberg
uncertainty relation, it shows nevertheless that even in the classical
realm there are 
situations where one can improve upon the venerated
averaging of $N$ independent measurement results by coupling the same
resources ``coherently'' to a $N+1$st system, and measuring the latter.  \\
 
{\em Model.}
Consider a classical harmonic oscillator with frequency $\omega_0\equiv\Omega$
harmonically coupled to $N$ other harmonic oscillators with frequency
$\omega_i$, 
$i=1,\ldots,N$. The Hamilton function (with masses $m_i=1$) reads 
\begin{equation} \label{}
H=\frac{1}{2}\sum_{i=0}^N\left(p_i^2+\omega_i^2q_i^2\right)+\frac{1}{2}\xi^2\sum_{i=1}^N\left(q_i-q_0\right)^2\,,
\end{equation}
where $p_i$ and $q_i$ are the canonical momenta and coordinates,
respectively, and $\xi^2$ denotes the coupling strength, such that
$\xi$ has 
the dimension of a frequency. This is
the parameter we want to 
determine. The
oscillators need not be mechanical oscillators, of course.  

The total potential energy in the problem can be rewritten as a quadratic
form, 
\begin{equation} \label{}
V(\bq)\equiv H-\sum_{i=0}^Np_i^2/2=\frac{1}{2}\bq^t\bC\bq\,,
\end{equation}
with $\bq^t=(q_0,\ldots,q_N)$, and 
\begin{equation} \label{}
\bC=\left(
\begin{array}{cccc}
\Omega^2+N\xi^2&-\xi^2&\ldots&-\xi^2\\
-\xi^2&\omega_1^2+\xi^2&0\ldots&0\\
\vdots&0&\ddots&\vdots\\
-\xi^2&0&\ldots&\omega_N^2+\xi^2  
\end{array}
\right)\,.
\end{equation}
The problem can be solved by diagonalizing $C$.  We are interested in
very small couplings,
$\xi^2\ll \omega_i^2$, $i=0,\ldots,N$, where diagonalization can be performed
perturbatively, starting from the uncoupled (squared) eigenfrequencies
$\lambda_i=\omega_i^2$, $i=0,\ldots,N$. Furthermore, we will restrict
ourselves to
the situation where the $\omega_i$ are narrowly distributed about a central
frequency $\overline{\omega}$, and sufficiently off-resonant from $\Omega$,
such that non-degenerate perturbation theory (PT) will suffice to obtain the
correction to $\Omega$. 
To order ${\cal O}(\xi^2)$ we have 
$\lambda_0=\Omega^2+N\xi^2
  +{\cal O}(\xi^4)$ and $
\lambda_l= \omega_l^2+\xi^2+{\cal
  O}(\xi^4),\,\,\,\,l=1,\ldots, N$.
The perturbed eigenmodes $\bu_l$ are summarized in the orthogonal
transformation matrix 
$\bU=(\bu_0,\bu_1\ldots,\bu_N)$ to order ${\cal O}(\xi^2)$ as
\begin{equation} \label{}
\bU=\left(\begin{array}{cccc}
1&-\frac{\xi^2}{\omega_1^2-\Omega^2}&\ldots&-\frac{\xi^2}{\omega_N^2-\Omega^2}
\\
\frac{\xi^2}{\omega_1^2-\Omega^2}&1&0&\ldots\\
\vdots&0&\ddots&0\\
\frac{\xi^2}{\omega_N^2-\Omega^2}&0&\ldots&1
\end{array}
\right)\,.
\end{equation}
We consider two sources of uncertainty: {\em i.)} an uncertainty in
the frequencies $\omega_i$, $i=1,\ldots,N$, and {\em ii.)} time
dependent noise.\\

{\em i.) Uncertainty in frequencies}. 
The equations of motion without external driving, $
\ddot{q}_i+\sum_j C_{ij}q_j=0
$
are decoupled into $N+1$ independent harmonic
oscillators by the transformation
$\bq\rightarrow \tilde{\bq}=\bU^t\bq$.  Solving them and transforming
back leads to the response of the central oscillator.  
In order to simplify expressions, we specialize to vanishing
initial speeds for all oscillators, $\dot{q}_j(0)=0$, $j=0,\ldots,N$.  We
will furthermore assume $N\xi^2\ll \Omega^2$, such that
$\sqrt{\lambda_0}=\Omega(1+N\xi^2/(2\Omega^2))+{\cal O}(\xi^4))$. This
limits $N$, but for very small $\xi^2$, $N$ can become very large (see
also the comments below for the validity beyond PT). To ${\cal
  O}(\xi^2)$ we then get 
\begin{eqnarray} \label{q0t}
q_0(t)&=&q_0(0)\cos((\Omega+\frac{N\xi^2}{2\Omega})t)
+\xi^2\sum_{j=1}^N\frac{q_j(0)}{\omega_j^2-\Omega^2}\\
&&\times\left(\cos((\Omega+\frac{N\xi^2}{2\Omega})t)
-\cos((\omega_j+\frac{\xi^2}{2\omega_j})t)\right)\,.\nonumber
\end{eqnarray}
The appearance of a phase shift that scales proportional to $N$ and the
parameter to be measured is reminiscent of phase superresolution
\cite{nagata_beating_2007}. This signal can be recovered by mixing the response
of the central oscillator with a $\cos(\Omega t)$ signal corresponding
to the 
unperturbed oscillator, i.e.~one multiplies $q_0(t)$ with
$\cos(\Omega t)$, which creates two signals, one that oscillates with the
sum of the two frequencies, the other with the difference. The latter varies
very slowly, and can be isolated with a low-pass filter. If we
assume the spectral width of the low-pass filter to be much smaller than
$2\Omega$ and $\omega_j-\Omega$, the remaining signal $s(t)$ reads
\begin{eqnarray}
s(t)&=&\left(q_0(0)+\xi^2r(\{\omega_i\})\right)\cos\left(\frac{N\xi^2}{2\Omega}t\right)\,\,\,,\nonumber
\\ 
\mbox{  where  } r(\{\omega_i\})&=&\sum_{j=1}^N\frac{q_j(0)}{\omega_j^2-\Omega^2}
\end{eqnarray}
is a random variable whose distribution is given by the distribution
$P(\{\omega_j\})$ of 
the $\omega_j$. 
The smallest uncertainty with which $\xi^2$ can be measured is given by
\cite{braunstein_statistical_1994} 
\begin{equation} \label{dximindef}
\delta\xi^2_{min}=\frac{\sigma(s(t))}{\sqrt{M}\left|\langle\frac{\partial
    s(t)}{\partial\xi^2}\rangle\right|}\,, 
\end{equation}
where $\langle\ldots\rangle$ means average over $P(\{\omega_j\})$, 
$\sigma(s(t))$ is the standard deviation of $s(t)$ with respect to this
distribution, and $M$ is the number of measurements. It has the
meaning of the smallest variation in $\xi^2$ that moves the average of
the signal at
least a distance given by the width of the distribution of the signal. 
A short calculation yields
\begin{equation} \label{}
\delta\xi^2_{min}=\frac{\xi^2\sigma(r)|\cos(\frac{N\xi^2}{2\Omega}t)|}{\sqrt{M}\left|
\frac{N}{2\Omega}t\left(q_0(0)+\xi^2\langle
r\rangle\right)\sin(\frac{N\xi^2}{2\Omega}t)-\langle r\rangle\cos\left(\frac{N\xi^2}{2\Omega}t\right)
\right|}\,.
\end{equation}
If one waits long enough ($N\xi^2t/(2\Omega)\gg 1$), the first term in the
denominator dominates.  If we set in addition $q_0(0)=0$, we obtain the final
result  
\begin{equation} \label{dxminfin}
\delta\xi^2_{min}=\frac{1}{N}\frac{2\Omega}{\sqrt{M}t}\left|\cot\left(\frac{N\xi^2t}{2\Omega}\right)\right|\,\frac{\sigma(r)}{\langle  
r\rangle} \,.
\end{equation}
The prefactor $1/N$ identifies this minimal uncertainty under the noise
process considered as ``Heisenberg-limited''.  Of course, this has
nothing to do with Heisenberg's uncertainty relation. Rather, 
we have considered a classical noise process (uncertainty in the
original frequencies $\omega_i$), but have found a way to reduce the
resulting smallest uncertainty with which the parameter $\xi^2$ can be
measured from a $1/\sqrt{N}$ scaling (that would be obtained by
measuring it separately from each system $i$ and the central
oscillator, and then averaging) to a $1/N$ scaling. The process how
this happens is completely analogous to the quantum mechanical collective
phase accumulation described in
\cite{braun_decoherence-enhanced_2009,braun_heisenberg-limited_2011}:
$N$ systems interact with a 
common central system, and lead to an accumulated phase proportional to $N$
and the parameter to be measured.  This manifests itself in an oscillation
with a frequency proportional to $N\xi^2$ in a corotating frame --- the
equivalent of homodyne detection in the quantum optical setting.  It also
leads to the same scaling with $t$, namely as $1/t$, and not the usual
$1/\sqrt{t}$.  It means that the sensitivity per square root of Hertz,
$\delta \xi^2_{min}\sqrt{t}$, still decreases as $1/\sqrt{t}$, just as in the
Heisenberg limited quantum case. Different from the quantum-mechanical case is,
however, the restriction to $N$ such that $N\xi^2\ll \Omega^2$. This
is true beyond the validity of PT, as is seen by analysing the very strong
coupling limit, where $\bC$ can again be diagonalized
analytically. Nevertheless, the method proposed here may be
advatageous for very small interactions, where $N$ can be very
large. 

{\em ii.) Time-dependent noise}
In the presence of time-dependent noise forces $f_i(t)$ acting on oscillator
$i$, the equations of motion in the original oscillator coordinates $q_i$
read 
\begin{equation} \label{}
\ddot{q}_i+\sum_j C_{ij}q_j=f_i(t)\,.
\end{equation}
After transformation to the eigenmodes $\tilde{q}_i$ defined through
$q_j=\sum_l U_{jl}\tilde{q}_l$ we get 
\begin{equation} \label{}
\ddot{\tilde{q}}_k+\lambda_k\tilde{q}_k=\sum_iU_{ki}^\dagger f_i(t)\equiv\tilde{f}_k(t)\,.
\end{equation}
A special solution of this equation can be
found with the help of the Greens-function of the harmonic oscillator. The back transformation to the original
coordinates gives for the central oscillator
\begin{eqnarray}
q_0(t)&=&q_0(0)\cos(\sqrt{\lambda_0}t)+\frac{\dot{q}_0(0)}{\sqrt{\lambda_0}}\sin(\sqrt{\lambda_0}t)\nonumber\\
&&+\int_0^t\frac{\sin\sqrt{\lambda_0}(t-t')}{\sqrt{\lambda_0}}f_0(t')dt'+{\cal O}(\xi^2)\,.
\end{eqnarray}
The noise enters here alread at order $\xi^0$, i.e.~perturbs even
the uncoupled central oscillator.  Nevertheless, we will now see that the
phase accumulation of the central oscillator due to the coupling to the $N$
other oscillators still leads to a $1/N$ scaling of the sensitivity.

We restrict ourselves again to $\dot{q}_j(0)=0$ for all $j=0,\ldots,N$ and 
$N\xi^2\ll \Omega^2$ such that
$\sqrt{\lambda_0}=\Omega(1+N\xi^2/(2\Omega^2))+{\cal O}(\xi^4))$, and
calculate the direct response to the noise. For simplicity we consider
noise with zero average, 
$\langle f_0(t)\rangle=0\,\,\forall t$, where $\langle\ldots\rangle$ means now
average over the noise-process. We then have $\langle
q_0(t)\rangle=q_0(0)\cos\sqrt{\lambda_0}t$ and
$\sigma^2(q_0(t))=\sigma^2(n(t))$, where 
\begin{equation} \label{}
n(t)=\int_0^t\frac{\sin\sqrt{\lambda_0}(t-t')}{\sqrt{\lambda_0}}f_0(t')dt'
\end{equation}
is the noise response.\\

{\em ii.a White noise}  Consider first white noise,
defined through $\langle f_0(t_1)f_0(t_2)\rangle=f_0^2T\delta(t_1-t_2)$,
where we have introduced a unit of time $T$ for dimensional grounds, in
addition to the force amplitudes $f_0$. One then immediately gets 
\begin{equation} \label{}
\sigma^2(q_0(t))=f_0^2T\int_0^t\frac{\sin^2\sqrt{\lambda_0}(t-t')}{\lambda_0}dt'\le \frac{f_0^2Tt}{\lambda_0}\,.
\end{equation}
In fact, for large times, $t\gg 1/\sqrt{\lambda_0}$, one has
$\sigma^2(q_0(t))\simeq \frac{f_0^2Tt}{2\lambda_0}$. All the dependence on $N$
of the sensitivity arises again from the derivative of $\langle q_0(t)\rangle$
and thus $\lambda_0$ with respect to $\xi^2$.  Inserting everything in
eq.(\ref{dximindef}), we are led to 
\begin{equation} \label{dximinN}
\delta\xi_{\rm min}^2\le
\frac{2f_0\sqrt{T/t}}{\sqrt{M}N|q_0(0)\sin(\sqrt{\lambda_0}t)|}\,, 
\end{equation}
where for large times still a factor $1/\sqrt{2}$ can be gained on the
rhs. Again, we 
see that 
the result scales as $1/N$.  The time dependence is different from the
previous case (and the typical quantum situation at the HL):
for large times the 
smallest resolvable $\delta\xi^2$ decays only as $1/\sqrt{t}$, just as in
the standard quantum limit.  \\

{\em ii.b Colored noise.} The above considerations are easily
generalized to colored noise.  In 
fact, unless the noise $f_0$ on oscillator 0 depends already at order $\xi^0$
on $N$ (which appears to be a highly artificial situation, since
without interaction the central oscillator should not ``know'' about the
number of additional oscilltors), $\sigma^2(n(t))$ is independent of $N$,
and the same scaling analysis concerning $N$ therefore applies and always
leads to a $1/N$ scaling of the sensitivity.  Only the time dependence will
differ. As a more 
general example, consider stationary colored noise with a correlation
function $\langle 
f_0(t_1)f_0(t_2)\rangle=f_0^2C(t_1-t_2)$ where we take $C(0)=1$, and
$C(-t)=C(t)$. One then easily finds the upper bound 
\begin{equation} \label{}
\sigma^2(q_0(t))\le \frac{2f_0^2}{\lambda_0}\int_0^tdt_-\,|C(t_-)|(t-t_-)\,.
\end{equation}
If the correlation function vanishes for $t>t_c$, one has 
$
\sigma^2(q_0(t))\le \frac{2f_0^2}{\lambda_0}b(t)
$
with 
\begin{equation} \label{}
b(t)=\left\{\begin{array}{ll}
tt_c-\frac{1}{2}t_c^2&t_c<t\\
\frac{1}{2}t^2 & t_c\ge t\,,
\end{array}\right.
\end{equation}
where we have used $|C(t)|\le |C(0)|$. 
Correspondingly, we have for the sensitivity
\begin{equation} \label{}
\delta\xi_{\rm min}^2\le \frac{2\sqrt{2} f_0}{\sqrt{M}}\frac{\sqrt{b(t)}}{Nt|q_0(0)\sin(\sqrt{\lambda_0}t)|}\,,
\end{equation}
which again scales as $1/N$.

A possible application of this ``coherent averaging'' technique might
be a novel way of measuring the gravitationl constant $G$, which is
one of the least well-determined natural constants with a 
relative 
uncertainty of order $10^{-4}$ and the 1986 CODATA recommended value
based on conflicting experimental results
\cite{gillies_newtonian_1997}.  One of 
the reasons for 
this dire situation is the extremely weak strength of the
gravitational interaction. This, and the impossibility to
shield the gravitational field from other disturbing bodies, render
the determination of the absolute value of $G$ very difficult, in
spite of continued strong interest, driven in part by
attempts to detect a variation of $G$ as function of distance, time,
or other physical quantities.
Since Cavendish's pioneering work in 1798,
essentially all lab-experiments attempting to measure $G$ were based
on a beam balance or a torsion pendulum, and measured either a static
response (some by counterbalancing deflections of the small test
masses), 
or a dynamic one (allowing frequency-specific analysis synchronized
with a 
periodic excitation, see e.g.~the recent attempt to measure deviations
of the $1/r^2$ behavior below the dark energy length scale of about
$85\mu$m 
\cite{Kapner07}). Our ``coherent averaging'' method suggests a new,
massively parallel way
of attempting to measure $G$ more precisely: Couple $N\gg 1$ torsion
balances graviationally to one central one
(consisting of $N$ further beams with masses fixed at the ends, and
all beams attached rigidly to a common 
axis).  Then measure the shift in frequency of that central oscillator
 as function of
$N$ and the positions of the test masses. This will enable a reduction
of 
the uncertainty in the measured value of $G$ with a scaling $1/N$
and should become competitive with traditional measurements for large
$N$. \\

{\em In summary,} the above analysis shows that even in the
classical realm there are  
situations where it is possible to beat the venerated procedure of
averaging $N$ measurement results of a
 physical quantity that 
leads to a $1/\sqrt{N}$ scaling of the sensitivity.  It can be
improved upon by coupling the 
$N$ 
samples ``coherently'' to a central oscillator that will pick up a
collective 
phase proportional to $N$ and the parameter to be measured, and yield
in the 
end a $1/N$ scaling with the number of samples available. Applications
might be found in the measurement of very weak interactions, such as
the gravitational interaction between lab-scale test masses,
suggesting a new, massively parallel way of determining the gravitational
constant.

\bibliography{bibs}

\end{document}